# Learning Historical and Chronological Time: Practical Applications

José Gómez Galán[a]

[a] *Director of CICIDE (Metropolitan University, AGMUS, Puerto Rico-United States & Catholic University of Avila, Spain).*
*Corresponding author: jogomez@suagm.edu / jose.gomez@ucavila.es*

**Abstract**
In the present article the necessity of introducing the teaching of time from the first educative stages is defended through a didactic process in which a feedback is produced between its chronological dimension (perception and measuring of physical time) and the historical (knowledge of time in the history of humanity). It has as its fundamental objectives the offering of pedagogical strategies which will contribute to the development of this competence in children, forming the basis of their future education as well as revising the classic educative theories about the learning of time. In particular, in relation to historical time where the employment of hyper-media timelines and the most original technology of improved virtual reality and 3D systems can contribute to an approach being given to their introduction for comprehension at an earlier age. Concurrently, the correct integration of those novel tools and telematic and computational systems in education, especially from the humanistic and analytical viewpoint, will permit the adoption of critical attitudes in the school with regard to the fictitious and virtual reality produced. It will contribute, in this manner, to facilitating a better comprehension of the modern world and the presence and functions that these play in society, in a constructivist context of the development of knowledge, science and culture[*].

## 1. Introduction: The Understanding of Time in Learning

The learning of chronological and historical time constitutes one of the bases of education in social sciences and humanities. Following the classic research of Piaget [1] [2] the understanding of chronological time must be the foundation of the understanding of historical time. As well, this author concludes that children have little conceptual understanding of time, which only begins to be properly understood from 11 years on and especially in adolescence [3]. Nevertheless, recent research, as will be shown later, has placed this in doubt and suggested that children at an earlier age, starting from five years old and even younger can comprehend historic time, understood as an ability to identify people, places, objects, clothing, etc, belonging to different periods of the past [4] [5]. These studies show that the development of chronological and historical understanding is more a process of learning than a process of cognitive development [2].

It is certain that, according to what was said previously, it is possible to deepen the development of the temporal notion from, especially, five years of age. The main means for developing this must be images, of course, and audiovisual media –owed moreover to its role in the technological world of today, as we have specified– is ideal for this purpose [6] [7]. Based on the studes of Fraisse [8], Brown [9], Friedman [10], Calvani [11], Vukelich & Thornton [12], Harnett [13], Gomez-Galan & Mateos [14], Barton [15], and Levstik & Barton [16], among others, we also argue that it is the right time for the beginning of the development of the sense of historical time. The characteristics of the society in which children develop makes it advisable. Today there is a need to train children from a relatively early age about everything they are forced to know and experience very early on, since it may condition their continuous formation in the future. We believe that not only the development of the notion of chronological time is to be found among them, but especially the characteristic of historical time. The excessive information offered by the audiovisual media (we could even say even a glut, and to which children are receptive) means the presence of continuous references to the historical, to *history*. What not too long ago was considered only a particular

---

and secondary area is now an element continuously present (because of the overload of information as has happened in so many other disciplines) in the life of the people.

To begin the introduction of historical time - of course as simply as possible-, in those moments, bearing in mind the proposal of Bruner´s *spiral curriculum* [17], will satisfactorily direct children, in the educational stages of elementary and secondary education, towards the specific and intricate meaning of the disciplinary concept of the same, which according to various studies, such as Jahoda [18], and subsequently developed by Pluckrose [19], is not achieved, until eleven years of age.

The foundation, for the beginning of the development of the sense of historical time will be, obviously, chronological. The main reasons for this is that the child, from the first moments of his life, will have experiences of change and sequences, both in his own body and in the environment, i.e. he will experience rhythms and routines that involve change and sequencing (my dog was born, was a puppy, grew, etc.; I get up, have breakfast, go to school, have a siesta, etc.), both fundamental elements for an approach to historical time. What we suggest is using the sense of chronological time that a child of this age has so that he approaches, not as a concept, of course, and for the reasons already given, that there was a time prior to the beginning of his life, that a changing world has always existed (which has taken on different forms) previous to their *current* world, which also evolved, and approach, thereby, a very basic temporal sequence of the *historical* past (without requiring the child to know of course, concepts like *change* or *sequence*) with the help of the basic chronological intervals which he experiences in his life and that can be determined.

The experiments carried out by Blyth [20] [21], also followed by Pluckrose [19], have shown that children of five and six years old have a certain prior understanding of what *antiquity* is. In our case we want to delve deep into this aspect, in such a way that children can have a first *contact* with the reality of the past of human beings (or the world), and its different stages, starting from the sense of chronological time. Images are crucial for this. He can discover and check through audiovisual (cartoons and movies) and visual media (picture books, comics, stories, cards, posters, slides, etc.), that long ago there were people who lived in different ways, with different customs, utensils and tools, clothes, buildings, animals and plants, etc., different scenes that corresponded to different times or periods of history: the world of dinosaurs, troglodytes, the Romans, the world of fairytales (castles, princesses, etc.), of Swordsmen and Musketeers, of pirates, Cowboys and Indians, etc. Teaching strategies may derive, in addition, from situating, on the basis of the sense of chronological time of a five-year-old child, these elements in time: dinosaurs are very, very *old*, much more than the Romans; the Romans are *older* than the pirates; the Musketeers are less *ancient* than the troglodytes; etc. (we say *old* here but any other expression can be used similar to them: *very old*, *long ago*, *a long time ago*, etc.).

In short, we can determine that it is possible, and necessary, to initiate an approach to the sense of historical time at this age, which will constitute thus the first pillars for further development in other educational stages. It is possible to develop different strategies and activities that will be highly appropriate for developing the sense of time among childrens in the second cycle of early childhood education (or elementary education, depending on educational structure of each country). In accordance with the above, we must first establish a set of essential activities focused on the development of chronological time, as a basis for the understanding of *historical time*. While it may be a parallel evolution, we believe that it is essential to follow this order, because the child, given its psycho-evolutive characteristics, experiences earlier to temporal experiences of an exclusively chronological type (biological rhythms, routines, etc.) which serve as a reference and cognitive support.

## 2. Strategies and Resources for the Development of the Sense of Chronological and Historical Time

Among the activities for developing the sense of chronological time, we can show several that we have developed in a school context with positive results, as we have discussed in another previous contributions [7] [14]. The most significant, which we list sequentially below, are as follows: (1) the teacher will talk to the children about the effects that come with the passage of time, so that these may begin to understand the idea that everything changes with the passing of it, with special emphasis on people, animals, plants, landscapes, seasons, etc. To do this, we can ask questions such as: do you remember how you were when you were small? What did you do then? How are you now? What did you play before that you do not play now? What will you be like in the future? What were your parents like when they were children? (2) Listen to, or visualize a story (in the medium preferred by the teacher) and, when finished, reconstruct the plot so the main ideas are explicit: what did the main character do first? What did he do after that? Why did he carry out those actions in that order? Would something have happened if he had not done what he did? In this way all of them can come to understand the temporal sequence of chronological time. Children who wish to do so can recount their previous experiences in this regard and the others they expect to have from this moment on, while the teacher can emphasize the ideas of *before and after*. (3) Play a traditional game that everyone knows, considered by the children as one of their favourites: the goose game. To do so, every 4 or 5 children will be given a board; in this way, in addition to other evident objectives such as using numerical series (in this case, the numbers 1, 2, 3, 4, 5 and 6, which are the ones shown on the dice) counting (the different squares), we will be *working* without any doubt, and, in addition, in a completely fun way, the notion of time, since the children will learn to wait their turn, to wait, to be aware that, as in other moments in life it is necessary to follow an order, they will understand better the terms *before*, *after*, *now*, *first*, *second*, *third*, *fourth*, etc., as they will be experiencing this. (4) Planting birdseed, beans, chickpeas, etc, in yogurt pots. They will take care of them and wait until the seedlings are born (waiting means being conscious of the passing of the days). Observing how plants are born and grow as time passes. (5) On a sheet of paper only showing a horizontal axis, children must begin situating the different events that have taken place in their lives: first, second, third, fourth and fifth birthdays, the birth of their little dog, the last Christmas day, etc. (6) Write on another sheet of paper, and also on a horizontal axis, some of the most relevant events that have taken place in their lives during the previous year: birthday, Easter holiday, summer vacation, back to school in the new term, etc.

All of these experiences will enable us to work with *chronological time* in children of this age. We should, naturally, lay the foundations of the sense of *historical time*. To initiate children in this other aspect of time, and in connection with the above, it will be convenient to start to do it through the personal history of each one: (7) For this purpose it will be ideal to use a family album, whereby the child can become aware of the existence of a past that is related to his life: he has parents (or has had) which were obviously born before he was born, in another time when many things were different (other cars, other apparel, other toys, etc.). This the children can check via different pictures.

On the basis of this we can establish more complex didactic activities that take into account the above: (8) The children can ask their parents to write their full names, their brothers too (the children´s uncles), the children of these (cousins) and their parents (grandparents) on a sheet of paper. The children can cut out these names and stick them on cardboard forming a small genealogical tree, helped by the clear indications of their teacher, while he comments on each of the generations. Of course he should emphasize the fact that every tree is different since it is something personal, the personal history of each one, and people are unique and unrepeatable. (9) Visualize every fifteen days, for example, an episode of the serie, *Once Upon a Time... Man* (1978, Albert Barillé, Studios Procidis; original version: *Il Était une Fois… l'Homme*), with the unique objective, in these moments, to continue opening the child´s mind to the possibility and certainty of a historical past. (10) The handling of illustrations (stories, comics, trading cards, posters, etc.) and in general material related with the cartoon *The Flintstones* (1960-1966, Hanna-Barbera Productions, Inc.) as well as with the film *The Land Before Time* (1988, Don Bluth, Universal Studios), and after having visualized some of it, we will establish three groups: some children will model with clay an animal or object which they have seen belongs to times past (such as a dinosaur, a Flintstones' car, etc.), others will model objects or animals from the present and a third group will materialize their fantasies about objects that will belong to people who will live many years in the future. The following day, when the figures are suitably dry, all the children will comment and compare them among themselves, to be able to later classify them according to different historical epochs. We will also make the most of this time (if the teacher considers it suitable and if the children have made reference to this objects), to tell them that as well as the fact that some of the figures that they have created do not exist in reality or will never exist (the ones they have invented from the future) other objects such as the Flintstones' car did not exist because cars, which are a later invention had not been invented yet. Or that dinosaurs lived a *long time* before people existed. All this data will not only be teaching that time is very extensive, and many things have happened during it, but at the same time it will show them that not everything that appears in the medium of communication, television (and by extension all current digital media) is *real* (for example, dinosaurs and people coexist only in *fiction*). So we will be establishing at the same time, which is so necessary today, the foundation for education about technology and current media.

Once we have put into effect these experiences, which have been useful for developing in the child both chronological time child development as well as giving him a very basic introduction to the sense of historical time, it will be possible to combine both of them and go a step further, and situate ourselves in a concrete way in the field of history. So, we propose the following activity: (11) Helped by a medium in which the child can locate different events of his life, which we introduce in experiences 5 and 6, the teacher will contrast another horizontal parallel axis with the one carried out by the pupils, where he will place different elements from different historical periods (e.g. troglodyte, a Roman, a medieval Lady, a steam engine, etc.). Then he will explain to the children that just like in his life, time goes on and makes him change (his birthday, the start of the school holidays, Christmas, etc.) the world also changes over *much*, *much*, *much time*, landscapes change, animals change, people change their wardrobe, the things they make, in inventions, etc. The use of these simple *timelines*, as a *metaphor* for the passage of time, will allow children to gain an initial notion of it, and it will be developed gradually in the following academic courses, according to Lynn [22], Haas [23], Hoodless [24], Masterman & Rogers [25], Alleman & Brophy [26], Gomez-Galan & Mateos [14], Prangsma [27], Prangsma, Van Boxtel & Kanselaar [28], Cox [29], Boyd-Davis [30], Sebba [31], Dixon & Hales [32], and Boyd-Davis, Bevan & Kudikov [33].

Of course, for the development of all these experiences we have, nowadays, more effective teaching resources due to their multimedia and hypermedia capabilities, much closer, moreover, to the world of the image in which children in these educational stages live, such as have shown Harnett [13], Hillis [34], Gomez-Galan & Mateos [14] Gomez-Galan [35], Wiley & Ash [36], Dilek, [37], Freeman & Philpott [38], Cox [29], Haydn [39], and Schoeman [40]. So, instead of establishing these timelines on sheets of paper or on the blackboard, for example, it is possible to do them on a computer, presented by a digital projector onto a screen, which will make all the elements that we introduce dynamic and interactive (photos or videos of the children, parents, holidays, etc., and also different selected historical items which can be drawings, photographs and fragments from cartoons, movies or video games, etc.). It would also be possible to use a digital whiteboard or various tablets, that would allow us to interact directly on the screen, etc. [7].

All these possibilities of the current technological society were not contemplated in studies conducted during the twentieth century about the introduction to the notion of historical time in children, and the alternatives that are opened up in this respect (as in many other educational dimensions) are immense. But they are also very dangerous, due to the very nature of this media, so they must be used with extreme care and following a rigorous pedagogical and didactic examination [41] [42]. It is necessary, therefore, that we stop to analyze it.

## 3. New Characteristics and Possibilities in the Current Educational Scene

Considering everything that we have discussed there should not be the least doubt that images are essential for the initiation and development of the sense of historical time. Without them it would be extremely difficult (if not impossible) to achieve this objective in the educational stages we are focusing on. However new technologies, whose backbone is precisely images, allow us to enhance their possibilities of application. We must take into account that, currently, audiovisual media and, in general, the media, modified by their use of recent computer and thematic technology, are the most powerful communication tools ever designed by human beings and that, in their totality, are called information and communication technologies (ICT). Due to its presence, quality and the power of images it is reaching heights that would have been unthinkable, and considered as science fiction a short time ago.

In relation to the problems we confront, we can say that they also have the capacity to be *windows to the past*. Consider, for example, a film such as *Jurassic Park* (1993, Steven Spielberg, Universal Studios), and its sequels and derivative productions. To visualize so effectively dinosaurs recreated using computer graphics is simply, as happens to the main characters in this film, to travel to the past, to contemplate in a surprisingly real way what these extinct animals were like. The same might be said, naturally, of any film set in historical times. What until recently had to be recreations based on models, costumes, film sets of papier-mâché, etc., nowadays is computer-generated with a quality and accuracy to what in historical science is really spectacular. In the more than twenty years that have elapsed since the info graphic possibilities of this film (i.e., the development of computer-generated moving images) the recreation of scenes of the past have multiplied exponentially, with the addition of the expansion of 3D Visual systems. Currently any film production of an historical nature allows us to contemplate scenes that reconstruct cities, monuments or landscapes of the past in an astonishing way (and everything we have stated can be applied to video games, that nowadays have budgets similar to or even higher than major movie productions).

In that sense, and reinforced by the use of this virtual images, today it is possible to have access to, as we said previously, genuine *windows to the past* children can look through in different ways: movies, web pages, computer programs, video games, etc. Never up to now have we had instruments of such quality to initiate children into and develop the sense of historical time. In addition, and as we said at the beginning of this work, it is essential to carry out an educational process from the first moment (naturally we recommend from early childhood education, and of course there is now no excuse, at no later than the first cycle of elementary education) that also helps to delimit fact from fiction in children of these ages.

All the guidelines and exercises that we have pointed out in the previous section can be developed using the new image technology in current audiovisual media (cinema, television, computer and Internet, video games, etc.) with more effective results because of the intrinsic quality they possess, which enables us to reconstruct the past in an extraordinary way. In addition we have the added motivation which they generate in children, since they are fully interactive tools (even cinematographic productions have their versions of video games that allow interaction) with an undeniably fun aspect since they are also part of their games and favourite toys.

Although there are some studies that have focused on exploring the possibilities of the current multimedia and hypermedia systems for teaching history, such as Cavoura, Tsaganou, Grigoriadou, Koutra & Samarakou [43], Wiley & Ash [36], Greene, Bolick & Robertson [44], Poitras, Lajoie & Hong [45], or Schoeman [40], there are hardly any contributions in this line about the development of the sense of historical time and, in particular, in these early educational stages, therefore it is a line open to the future.

However, and having these sophisticated iconic means, we cannot confine ourselves to the above proposals. This is due to the fact that the newest ICT allows the creation of specific tools that can contribute to the achievement of this objective in a special way. Because of limitations of space we will only focus on one that will serve as an example of their real possibilities, not only in the development of the sense of historical time, but in virtually all of the processes of teaching and didactics of history.

The proposal we are going to make is the creation of multidimensional and multimedia *timelines*. This resource has been used since long ago in their printed versions. They are very useful for sorting and classifying historical events using the metaphor of a spatial line that corresponds with temporal becoming, in such a way that it is possible to establish the *before* and *after* of a particular point (event or historical fact). In addition, the present moment would appear at the end of that line through they are viewing the concept of the evolution of history that flows into today.

Of course timelines, in their traditional printed form, would be a teaching resource that we could consider appropriate to use but not before secondary school education, except perhaps in the last cycle of elementary education, and in this particular circumstance using a multitude of graphic elements that can help aid interpretation. However with the use of new information technologies and telematics it is possible to create timelines with multimedia features, with iconic elements of all kinds, videos (with fragments of films, specific info graphic productions, elements of video games, etc.), possibilities for interaction, presented by projectors, and a long etcetera. In these multidimensional and multimedia timelines it would be possible to not only include facts and historical events but all kinds of processes, illustrations of an era, social life, cultural material, recreations of amazing quality and perfection, etc., and in conjunction with full interactivity in such way that the student can move freely through it and, in the context of the metaphor establish connections between chronological and historical time. In this sense, these multimedia and multidimensional time lines could be used with much younger age groups, even with children of five years of age in early childhood education and, of course, in the first cycle of elementary education, naturally it would be adapted to their cognitive abilities and using all the elements that we have discussed in the previous points. It would be an example of the possibilities offered by current audiovisual media, developed through the latest computer and telematic advances for the teaching of historical time and, in general, of history.

## 4. Virtual Reality and Historical Time Learning

But the potential as vast and the development of new technologies will create applications that we cannot imagine. For example, the recent description offered by Novak & Canas [46] of the construction of knowledge through virtual means with the development of conceptual maps that could be applied to the learning of History. Much closer to our goal would be contributions from Eden [47] and Eden & Passing [48] describing the possibilities of 3D virtual reality for the perception of sequential time in applications within the context of special educational needs (SEN), experiences that could be taken into account and used in a teaching context, for example, early childhood education. In our case, we have already worked on the educational development of virtual reality [49] [50], we can establish that, without a doubt, their applications at the beginning of the development of the notion of historical time in children could be very interesting, although we do not believe that they can get better results than current technology being most widely employed.

Virtual reality through the 3D scenes that it generates and the possibilities of interaction offered in this three-dimensional environment, will no doubt have a first class educational application for the teaching of history, but not especially for an initiation into the notion of historical time (although there are several interesting research, e.g., Foreman, Boyd-Davis, Moar, Korallo, & Chappell [51], Korallo [52], and Korallo, Foreman, Boyd-Davis, Moar & Coulson [53]). Which is not to say that the future will not hold, as we have indicated, sophisticated but as yet unknown technologies suitable for this purpose.

We must bear in mind that audiovisual media, and within this concept are included all those offered currently by the digital paradigm –with images being the most important– provides information of a quality not possible before, and of a volume that was unthinkable just a few years ago. The influence they have in our society, and consequently on education, means that they must not only be considered as teaching resources that can be extremely useful to develop didactic processes, in the field of the social sciences/humanities and, in particular, history, but they must be considered in the school as fundamental elements of our world that need to be studied and analyzed.

An example of this is, naturally, the virtual reality which we have referred to. If we said previously that current audiovisual media are *windows to the past* due to the quality and quantity of the iconic information provided, the extension of this technology used for generating scenarios in 3D will allow students in early childhood and elementary school education in the future to, practically speaking, perform *virtual time travel to the past*. It will be the most effective means for the creation of a *reality* as it is, or has literally been, i.e. history.

We find ourselves before a situation, taking into account the age of these children, in which only using texts or simple images as a teaching resource would be at an absolute disadvantage compared to this three-dimensional audiovisual environment. Even traditional media would find itself far behind the possibilities of this technology. Not only would it be possible for a student to view the world of dinosaurs, or contemplate a city from the middle ages (if they are used for the proposed examples, traditional media such as film, television, or video), but rather it would be possible to *walk* through the Triassic jungle or through the narrow streets of the different guilds of a town interacting with the environment and, therefore, accessing knowledge in a process of discovery. Students (equipped with stereoscopic glasses and touch gloves, or new technological systems as yet unknown) would have the option of *moving* in these virtual spaces which recreate an *historical reality*, accessing knowledge about it.

If in addition virtual reality develops further, as it will do, the Internet will offer a new dimension in the way of communicating and the access to information, proving to be ideal for the development of meaningful learning [41] [42], in a constructivist context [54]. Although this, of course, will involve objectives beyond the development of the notion of historical and chronological time.

## 5. Discussion and Conclusions

Best practice for learning is changing. Digital technologies enable learning and teaching to have more possibilities and opportunities. There has never been so much at our disposal, from the didactic point of view, with which to teach historical and chronological time, to teach, ultimately, history and its chronology. Now, it is not just about markers, murals, textbooks and slides, or the screening of some historic film in the classroom. Nor does it amount to working with *timelines* in cardboard with limited scope for interaction and interpretation. In a not too distant future, one a lot closer than most of us realize, students could leap from one epoch to the next in a virtual time *tunnel* in which a visit to the Rome of Emperor Trajan, could be followed by one to the Paris of Louis XIV or by the simple movement of a lever the child could go back (in a time machine operated by virtual reality systems that may well be designed like those envisaged by H.G. Wells) to visit shamans in Solutrean, who are painting magnificent bisons on the cave walls of Altamira. And even more important than all of that is that they would be able to interact with all those elements (in this last case, for example, they could even join the shamans in their painting or move objects around that they find there, like a bifacial *laurel leaf*). They will be able to access knowledge of *historical reality* that has never been produced before. Of course making that historic fact understood, its consequences and the causes, the economic, political or social processes that influence the evolution of history will continue to be an essential function of the teacher, but the student's access to the *past*, to the *time*, to the essence of history will be augmented exponentially.

Nevertheless, and concurrently to that, the employment of this technology will present important problems that will need to be addressed through rigorous examination and thorough preparation in a pedagogical context. For example, students could have difficulties distinguishing fiction from reality, especially in early ages. This could perhaps be avoided by simply not using these new mediums and resources, but that would be impossible because these tools form part of everyday life. In this hyper-technological and media-orientated society in which we live, the ICT are not didactic tools that have been designed for the variety of purposes that they can now be used for. Their principle applications today are in the leisure and entertainment fields and their prime use for infants and youth in general is in the fields of video games, television and internet in general. They represent a unique medium for multiple messaging, the product of techno-media convergence [14] [35]. The teacher therefore in the classroom, is obliged to create critical attitudes on the part of students towards the influential power of these technologies, which tend all the more towards consumerism, and lead to a variety of different addictions and dependencies, and hold the lives of many people in a *virtual*, rather than the *real* world. This will drive new educative processes to set adequate guidelines in the understanding of a society in which these activities will be absolute. It is more than the consideration of didactic resources, it implies a fundamental study of the education of future citizens.

Therefore, and taking this perspective into account, a study of the didactic advantages must be undertaken and at the same time an analysis and critical study carried out. And as is being shown in this current study, the possibilities of contributing to the learning of chronological and historical time are important. But also, and precisely because of the presence of ICT in every field of life and its capacity to offer new and different stimuli, it should be understood as an essential objective of instruction that children from an early age will be exposed to this technology and education must play a role in the training of the student body for its correct management and for the betterment of its comprehension. All of this supports the foundations of an adequate education in the field of social science and humanities where the concept of *time* is essential.

**References**

[1] Piaget, J. (1946). *Le Développement de la Notion de Temps chez l'Enfant*. Paris: Presses Universitaires de France.
[2] Piaget, J. (1965). *The Child's Conception of the World*. Totowa: Littlefield Adams.
[3] De Groot-Reuvekamp, M. J., Van Boxtel, C., Ros, A., & Harnett, P. (2014). The Understanding of Historical Time in the Primary History Curriculum in England and the Netherlands. *Journal of Curriculum Studies*, 46 (4), 487-514.
[4] Thornton, S. J., & Vukelich, R. (1988). Effects of Children's Understanding of Time Concepts on Historical Understanding. *Theory & Research in Social Education*, 16 (1), 69-82.
[5] Hinde, E. R., & Perry, N. (2007). Elementary Teachers' Application of Jean Piaget's Theories of Cognitive Development during Social Studies Curriculum Debates in Arizona. *The Elementary School Journal*, 108 (1), 63-79.
[6] Gomez-Galan, J. & Mateos, S. (2003). La Enseñanza del Tiempo Histórico en la Educación Infantil. In J.L. Geiger (ed.), *Proceedings of First South American History Congress*. Santa Cruz de la Sierra (Bolivia): Bolivian Studies Association.
[7] Gomez-Galan, J. (2009). Empleo de la Imagen mediante Medios Audiovisuales en la Enseñanza del Tiempo Histórico: Posibilidades en el Ámbito de Educación Infantil y Primaria. *Elvas/Caia. Revista Internacional de Cultura e Ciencia*, 7(1), 245-271.
[8] Fraisse, P. (1957). *Psychologie du Temps*. Paris, P.U.F.
[9] Brown, A. L. (1975). Recognition, Reconstruction, and Recall of Narrative Sequences by Preoperational Children. *Child Development*, 156-166.
[10] Friedman, W. J. (1978). Development of Time Concepts in Children. *Advances in Child Development and Behavior*, 12, 267-298.
[11] Calvani, A. (1988). *Il Bambino, il Tempo, la Storia*. Firenze: La Nuova Italia.
[12] Vukelich, R., & Thornton, S. J. (1990). Children's Understanding of Historical Time: Implications for Instruction. *Childhood Education*, 67 (1), 22-25.
[13] Harnett, P. (1993). Identifying Progression in Children's Understanding: The Use of Visual Materials to Assess Primary School Children's Learning in History. *Cambridge Journal of Education*, 23, 137-154.
[14] Gomez-Galan, J. & Mateos, S. (2004). Pautas para el Inicio del Desarrollo del Sentido del Tiempo Histórico en el Niño de Segundo Ciclo de Educación Infantil. En M. D. García Fernández & V. Marín Díaz (ed.). *La Educación Infantil y la Formación del Profesorado hacia el Siglo XXI: Integración e Identidad* (pp. 477-485). Córdoba: Universidad de Córdoba.
[15] Barton, K. C. (2011). Enriching Young Childrens Understanding of Time. *Primary History*, 59, 16-18.
[16] Levstik, L. S. & Barton, K. C. (2015). *Doing History: Investigating with Children in Elementary and Middle Schools*. Fifth edition. New York: Routledge.
[17] Bruner, J. S. (1960). *The Process of Education*. New York: Harvard University Press.
[18] Jahoda, G. (1963). Children's Concepts of Time and History. *Educational Review*, 15 (2), 87-107.
[19] Pluckrose, H. (1993). *Children Learning History*. Hertz: Simon & Schuster.
[20] Blyth. J. (1984). *Place and Time with Children Five to Nine*. London: Croom Helm.
[21] Blyth, J. (1994). History: 5 to 11. London: Hodder & Stoughton.
[22] Lynn, S. (1993). Children Reading Pictures: History Visuals at Key Stages 1 and 2. *Education*, 21 (3), 23-29.
[23] Haas, M. (2000). A Street through Time Used with Powerful Instructional Strategies. *Social Studies and the Young Learner,* 13(2), 20-23.
[24] Hoodless, P. A. (2002). An Investigation into Children's Developing Awareness of Time and Chronology in Story. *Journal of Curriculum Studies*, 34 (2), 173-200.

[25] Masterman, E., & Rogers, Y. (2002). A Framework for Designing Interactive Multimedia to Scaffold Young Children's Understanding of Historical Chronology. *Instructional Science*, 30 (3), 221-241.
[26] Alleman, J., & Brophy, J. (2003). History is Alive: Teaching Young Children About Changes over Time. *Social Studies*, 94 (3), 107-111.
[27] Prangsma, M.E. (2007). *Multimodal Representations in Collaborative History Learning*. Unpublished Ph.D. thesis, Utrech: Utrecht University
[28] Prangsma, M. E., Van Boxtel, C. A., & Kanselaar, G. (2008). Developing a 'Big Picture'. Effects of Collaborative Construction of Multimodal Representations in History. *Instructional Science*, 36 (2), 117-136.
[29] Cox, C. (2012). *Literature Based Teaching in the Content Areas*. Thousand Oaks: SAGE Publications.
[30] Boyd Davis, S. (2012). History on the Line: Time as Dimension. *Design Issues*, 28 (4), 4–17.
[31] Sebba, J. (2013). *History for All*. London: Routledge.
[32] Dixon, L., & Hales, A. (2013). *Bringing History Alive Through Local People and Places: a Guide for Primary School Teachers*. London: Routledge.
[33] Boyd Davis, S, Bevan, E. & Kudikov, A. (2013). Just in Time: Defining Historical Chronographics. In J. P. Bowen & S. Keene. *Electronic Visualisation in Arts and Culture* (pp. 243-257). London: Springer
[34] Hillis, P. (2002). Multi-media and History Education: A Partnership to Enhance Teaching and Learning. *Educational Media International*, 39 (3-4), 307-315.
[35] Gomez-Galan, J. (2007). Los Medios de Comunicación en la Convergencia Tecnológica: Perspectiva Educativa. *Comunicación y Pedagogía: Nuevas Tecnologías y Recursos Didácticos*, 221, 44-50.
[36] Wiley, J. & Ash, I. K. (2005). Multimedia Learning of History. In R.E. Mayer. *Cambridge Handbook of Multimedia Learning* (pp. 375-391). New York: Cambridge University Press.
[37] Dilek, D. (2009). The Reconstruction of the Past through Images: An Iconographic Analysis on the Historical Imagination Usage Skills of Primary School Pupils. *Educational Sciences: Theory and Practice*, 9 (2), 665-689.
[38] Freeman, J. & Philpott, J. (2009). Assessing Pupil Progress: Transforming Teacher Assessment in Key Stage 3 History. *Teaching History*, 137, 4-12.
[39] Haydn, T. (2012). Information and Communications Technologies in the History Classroom. In J. Arthur, & R. Phillips (Ed.). *Issues in History Teaching*. Second Edition. (pp. 98-113). London: Routledge.
[40] Schoeman, S. (2013). Presentation Technology as a Mediator of Learners' Retention and Comprehension in a History Classroom. *Yesterday and Today*, (9), 67-90.
[41] Gomez-Galan, J. (2015). Media Education as Theoretical and Practical Paradigm for Digital Literacy: An Interdisciplinary Analysis. *European Journal of Science and Theology*, 11 (3), 135-150.
[42] Gómez Galán, J. & Mateos, S. (2002). Retos Educativos en la Sociedad de la Información y la Comunicación. *Revista Latinoamericana de Tecnología Educativa*, 1 (1), 9-23.
[43] Cavoura, T., Tsaganou, G., Grigoriadou, M., Koutra, D. & Samarakou, M. (2005). Learning History in the Constructive Hypermedia Environment of Kastalia. En P. Kommers y G. Richards (Eds.), *Proceedings of World Conference on Educational Multimedia, Hypermedia and Telecommunications 2005* (pp. 2389-2390). Chesapeake, VA: AACE.
[44] Greene, J. A., Bolick, C. M., & Robertson, J. (2010). Fostering Historical Knowledge and Thinking Skills using Hypermedia Learning Environments: The Role of Self-Regulated Learning. *Computers & Education*, 54, 230-243.
[45] Poitras, E., Lajoie, S., Hong, Y.J. (2011) The Design of Technology-Rich Learning Environments as Metacognitive Tools in History Education. *Instructional Science*, 40 (6), 1033-1061.
[46] Novak, J. D. & Canas, A. J. (2008). *The Theory Underling Concept Maps and How Construct Them*. Pensacola: Florida Institute for Human and Machine Cognition.
[47] Eden, S. (2008). The Effect of 3D Virtual Reality on Secuencial Time Perception among Deaf and Hard-of Hearing Children. *European Journal of Special Needs Education*, 23 (4), 349-363.
[48] Eden, S. & Passing, D. (2007). Three-dimensionality as an Effective Mode of Representation for Expressing Sequential Time Perception. *Journal of Educational Computing Research*, 37 (1), 51-63.
[49] Gomez-Galan, J. (2001). Aplicaciones Didácticas y Educativas de las Tecnologías RIV (Realidad Infovirtual) en Entornos Telemáticos. In C. Preciado (ed.). *XIII Congreso Internacional de Ingeniería Gráfica: Eliminando Fronteras entre lo Real y lo Virtual*, Badajoz: AEIA
[50] Gomez-Galan, J. (2002). Education and Virtual Reality. En N. Mastorakis (ed.). *Advances in Systems Engineering, Signal Processing and Communications*. (pp. 345-350). New York: WSEAS Press.

[51] Foreman, N., Boyd-Davis, S., Moar, M., Korallo, L., & Chappell, E. (2008). Can Virtual Environments enhance the Learning of Historical Chronology? *Instructional Science*, 36 (2), 155-173.
[52] Korallo, L. (2010). *Use of Virtual Reality Environments to Improve the Learning of Historical Chronology*, Ph.D. thesis, London: Middlesex University.
[53] Korallo, L., Foreman, N., Boyd-Davis, S., Moar, M., & Coulson, M. (2012). Do Challenge, Task Experience or Computer Familiarity Influence the Learning of Historical Chronology from Virtual Environments in 8–9 Year Old Children? *Computers & Education*, 58(4), 1106-1116.
[54] Ausubel, D. P. (2000). *The Acquisition and Retention of Knowledge: a Cognitive View*, Boston: Kluwer Academic Publishers.